\documentclass[review]{elsarticle}

\usepackage[utf8]{inputenc}
\usepackage{lineno,hyperref}
\modulolinenumbers[5]
\usepackage{amsmath}
\usepackage{amsfonts}
\usepackage{graphicx}
\numberwithin{equation}{section}
\usepackage{mathtools}
\DeclarePairedDelimiter\abs{\lvert}{\rvert}

\journal{Journal of \LaTeX\ Templates}

\begin{document}

\begin{frontmatter}

\title{Open string tachyon is the stringy inflaton}

\author{Hongsu Kim\fnref{myfootnote}}
\address{Center for Theoretical Astronomy, Korea Astronomy and Space Science Institute, Daejeon 34055, Republic of Korea}
\fntext[myfootnote]{chris@kasi.re.kr}

\begin{abstract}
Regarding the current status of the superstring theory, despite its fully developed contents of all interactions in nature, it is no doubt that unfortunately it lacks a well established early universe cosmology content, that is, the inflationary universe. In the present work, we would like to address precisely this issue. To be more concrete, we would like to demonstrate that remarkably it is the well established open string tachyon condensation which is essentially equivalent to the stringy inflation that has been missing and hence long-seeking by nearly the entire string community.
\end{abstract}

\begin{keyword}
superstring, inflationary universe, open string, tachyon 
\end{keyword}

\end{frontmatter}

\section{Introduction}
Ever since the advent of the first string revolution up until the recent development of the second string revolution, the most disturbing feature of the present superstring theory is that the theory lacks the well established event in the early universe cosmology, that is, the inflationary universe. Put differently, despite the long-seeking effort of the string community, the stringy inflaton that is supposed to drive the early universe inflation has been missing \cite{1}. In the present work, however, we would like to demonstrate that remarkably it is the well established open string tachyon condensation which, upon coupling to the gravitation, is essentially equivalent to the stringy inflation. That is, unlike one's plain conjecture, the unwelcome tachyon in open string theory turns out to be the long-seeking, mysteriously missing stringy inflaton. Being one of the best candidates for theory of quantum gravity as well as the unification of all the four forces in nature, the current status of the superstring theory is expected to involve well established features in the early universe, most notably, the inflationary universe paradigm. The forefront community of the superstring theory, therefore, has been extensively searching for this early universe inflation content in the entire context of the superstring theory. Unfortunately, however, despite the careful and extensive search, no such qualified candidate for stringy inflaton scalar field has been discovered thus far. This is, indeed, quite a frustrating state of the affair. To be established as a fully qualified theory of everything plus theory of quantum gravitation, finishing up the superstring theory with a relevant stringy inflaton scalar field candidate is indeed the pressing objective of the current superstring community. Along this line, early universe cosmology community of the superstring theory worldwide has been working hard toward this objective. And among such hard, extensive efforts, the most remarkable and hence the best known challenges so far are those of KKLT, and references therein \cite{2} and KKLMMT, and references therein \cite{3}.  These challenges, however, involve rather artificial construction of stringy inflaton by manipulating it in the pool of string moduli fields. Unfortunately, it is fair to say, therefore, that up until now, there is no successful manipulation of the best qualified stringy inflaton drawing the entire string community's consensus. With this state of the affair, we now would like to exhibit that we finally have discovered such a long-seeking stringy inflaton scalar field that has been hiding deeply from the intensive searches. To be more concrete, in the present work we would like to demonstrate that amazingly, it is the open string tachyon which is the fully qualified stringy inflaton. Put differently, it is the open string tachyon condensation process that is the long-seeking stringy inflation "in disguise". After all, it is no wonder why we have had such a long hard time finding out the stringy inflaton as it has been hiding way deep inside.

Along this line, it is quite promising that the supergravity analogue of the tachyon potential that the present author explored and published sometime ago \cite{4} can serve as a successful candidate for such a flat stringy inflaton potential as it possesses purely logarithmic dependence on the tachyon-field and hence is slowly-varying.

\section{Stringy inflation accompanying the open string tachyon condensation}
In the present work, we would like to demonstrate for the first time that a successful  stringy inflation accompanies the (open string) tachyon condensation in the unstable, say,  brane($D6$)-antibrane($\bar{D}6$) system. One might wonder why we particularly consider ($D6-\bar{D}6$) system in addressing the issue of open string tachyon condensation. As we will take this set-up throughout this work, perhaps a reasonable rationale might be necessary to render our present work convincingly natural. Such rationale of ours goes as follows. From the well established table for intersecting $D$ brane systems, it is known for some time that the ($D6-\bar{D}6$) system is one of the unstable, non-BPS brane systems. Next, under the spirit of Sen's conjecture, the non-BPS, unstable brane systems get stabilized  upon open string tachyon condensation on it.  This is the motivation for our set-up; open string tachyon condensation on the ($D6-\bar{D}6$) system.

The background which motivates this attempt of ours is as follows; Recently, motivated by a series of suggestions made by A. Sen, some intriguing possibility of tachyonic inflation on unstable, non-BPS branes has been proposed and explored. And there, "the tachyon field", whose dynamics on a non-BPS brane is described by an extended version of Dirac-Born-Infeld (DBI) and Chern-Simons (CS) action terms contain both the kinetic and potential energy terms. To be more concrete, we exhibit in the present work that the rolling tachyon field, plays the role of "stringy inflaton" driving an inflationary phase on the brane world.

And in order for this tachyonic inflation scenario on unstable, non-BPS branes to work and hence is valid, say, in the context of new inflation-type model, the most crucial condition for the slow roll-over regime to take place is the nearly-flat geometry of the tachyon potential. Tachyon potential with such highly restricted condition/behavior has finally been constructed by the current author \cite{4}. In order eventually to address the issue of tachyonic inflation scenario, therefore, we first begin with such tachyon potential constructed in \cite{4}.

\section{Supergravity analogue of the tachyon potential in the unstable brane$(D6)$ - antibrane$(\bar{D}6)$ system}
\subsection{In the absence of the magnetic field (eq.(3.10) of \cite{4})}
\begin{align}
V(T^*T)&=V(\abs{T}^2)\\
&=-2m\left[l^{-1}_s\left(\ln\frac{\abs{T_0}^2}{\abs{T}^2}\right)^{-1/2}\left\{m+\left(m^2+l^2_s\ln\frac{\abs{T_0}^2}{\abs{T}^2}\right)^{1/2}\right\}-1\right]
\end{align}
where $"m"$ denotes the ADM-mass of a single $D6$-brane and $"l_s"$ denotes the string scale, which is of order the Planck scale.

Note that in order for this supergravity description to be valid, we should demand in this expression for the tachyon potential, $m^2\ge l^2_s$. Next note that at the local vacuum $T=0$, $V(\abs{T}^2)=0$ consistently with the expectation. At the global vacuum $T=T_0$, however, $V(\abs{T_0})\rightarrow\-\infty$, namely the minimum of the tachyon potential constructed from this supergravity description is unbounded below. This is rather discouraging but it is interesting to note that by introducing an appropriate magnetic field (i.e., RR F7-brane) aligned with the axis joining the brane - antibrane pair with a proper strength $B$ removing all the conical singularities of the $(D6-\bar{D}6)$ solution, this tachyon potential can be regularized, i.e., can be made to be bounded from below. Thus we now turn to this case.
\paragraph{More on the "magnetic field" nature}
The supergravity solution representing unstable $(D6-\bar{D}6)$ pair system possesses inherent conical singularities. They, however, can be eliminated by intersecting ($D6-\bar{D}6$) system in parallel with the RR F7-brane with proper strength $B$.
And interestingly enough, (in parallel) intersecting this RR F7-brane plays the role of regularizing the unbounded and thus runaway nature of the supergravity analogue of the (open string) tachyon potential in the $(D6-\bar{D}6)$ system.

\subsection{In the presence of the magnetic field (eq.(3.17) of \cite{4})}
\begin{align}
V(T^*T)&=V(\abs{T}^2)\\
&=\frac{-2m^2}{\left[m^2+l^2_s\ln\frac{\abs{T_0}^2}{\abs{T}^2}\right]^{1/2}}=-2m^2\left[m^2-l^2_s\ln\frac{\abs{T}^2}{\abs{T_0}^2}\right]^{-1/2}.
\end{align}

\section{Open string tachyon condensation followed by the stringy inflation}
Now, consider the self-gravitating, complex, inflation scalar field(inflaton) $\phi(x)$, which, in the present work, is taken to be the (open string) tachyon field $T(x)\equiv\phi(x)$, described by the total action given by; 
\begin{equation}
S=S_G+S_T
\end{equation}
where
\begin{align}
S_G&=\frac{1}{16\pi G}\int d^4x\sqrt{g}R,\\
S_T&=\int d^4x\sqrt{g}\left[-g^{\mu\nu}(\partial_\mu T^*)(\partial_\nu T)-V(T^*T)\right].
\end{align}
Next, as a routine process, by extremizing this total action with respect to $g_{\mu\nu}$ and $(T^*, T)$, one gets the Einstein field equation and the classical, Euler-Lagrange's equation of motion for the tachyon field, respectively. We thus start with Einstein field equation which results upon extremizing this total action with respect to the metric $g_{\mu\nu}$;
\begin{align}
R_{\mu\nu}-\frac{1}{2}g_{\mu\nu}R&=8\pi GT^T_{\mu\nu},\\
T^T_{\mu\nu}&=\frac{2}{\sqrt{g}}\frac{\delta S_T}{\delta g^{\mu\nu}}\\
&=2(\partial_\mu T^*)(\partial_\nu T)-g_{\mu\nu}\left[g^{\alpha\beta}(\partial_\alpha T^*)(\partial_\beta T)+V(T^*T)\right]
\end{align}
where we used
\begin{align}
\delta S_T&=\frac{1}{2}\sqrt{g}T^{\mu\nu}_T8g_{\mu\nu},\\
\delta\sqrt{g}&=\frac{1}{2}\sqrt{g}g^{\alpha\beta}\delta g_{\alpha\beta},\\
\delta(\sqrt{g}g^{\mu\nu})&=-\sqrt{g}\left[g^{\mu\alpha}g^{\nu\beta}-\frac{1}{2}g^{\mu\nu}g^{\alpha\beta}\right]\delta g_{\alpha\beta}.
\end{align} 
And as usual, throughout this work, we shall assume that the universe is homogeneous and isotropic described by the FRW-metric
\begin{align}
ds^2&=g_{\mu\nu}dx^\mu dx^\nu\\
&=-dt^2+a^2(t)\left[\frac{dr^2}{1-kr^2}+r^2(d\theta^2+\sin^2\theta d\phi^2)\right]
\end{align}
then the tachyon field, i.e., complex inflaton scalar field in this homogeneous and isotropic background spacetime should be so as well. Then the matter energy density, namely the energy density of the complex inflaton field, i.e., the tachyon field is given by;
\begin{align}
\rho^T&=T^{00}_T=\dot{T}^*\dot{T}+(\nabla T^*)\cdot(\nabla T)+V(T^*T)\\
&=\abs{\dot{T}}^2+V(T^*T)
\end{align}
and the time-time component of the Einstein field equation, known as the Friedmann equation is given by;
\begin{equation}
\left(\frac{\dot{a}}{a}\right)^2+\frac{k}{a^2}=\frac{8\pi G}{3}\rho^T
\end{equation}
from which we can read off the Hubble parameter by neglecting the term $\frac{k}{a^2}$ which would vanish due to the occurrence of inflaton, that is,
\begin{equation}
H=\left(\frac{\dot{a}}{a}\right)=\sqrt{\frac{8\pi G}{3}\rho^T}
\end{equation}
of which the solution is obviously 
\begin{equation}
a(t)=e^{H(t)}
\end{equation}
where $H(t)$ is given by eq.(4.15). Even at this stage, therefore, it is obvious that the presence of open string tachyon, namely, the stringy inflaton renders the spacetime inflation to take place.

Next, we turn to the classical, Euler-Lagrange's equation of motion for the complex inflaton scalar field, i.e., the tachyon field $T(x)$ which results upon extremizing the total action given above with respect to the complex tachyon fields $T^*(x)$ and $T(x)$;
\begin{align} 
\begin{cases}
\frac{\delta S_T}{\delta T^*}-\nabla_\mu\frac{\delta S_T}{\delta(\nabla_\mu T^*)}&=0\\
\sqrt{g}\left[\nabla_\mu\nabla^\mu T-\frac{\delta V}{\delta T^*}\right]&=0
\end{cases}\\
\begin{cases}
\frac{\delta S_T}{\delta T}-\nabla_\mu\frac{\delta S_T}{\delta(\nabla_\mu T)}&=0\\
\sqrt{g}\left[\nabla_\mu\nabla^\mu T^*-\frac{\delta V}{\delta T}\right]&=0
\end{cases}
\end{align}
At this point, using;
\begin{align}
\nabla_\mu\nabla^\mu T&=\square T=\frac{1}{\sqrt{g}}\partial_\mu(\sqrt{g}g^{\mu\nu}\partial_\nu T)\\
&=\frac{1}{\sqrt{g}}(\partial_\mu\sqrt{g})g^{\mu\nu}(\partial_\nu T)+g^{\mu\nu}\partial_\mu\partial_\nu T\\
&=\Gamma^\lambda_{\lambda\mu}g^{\mu\nu}(\partial_\nu T)+g^{\mu\nu}\partial_\mu\partial_\nu T
\end{align}
and further using; $T(x)=T(t)$
\begin{equation}
\Gamma^\lambda_{\lambda t}g^{tt}(\partial_tT)=-\ddot{T}=-3\left(\frac{\dot{a}}{a}\right)\dot{T}=-3H\dot{T}.
\end{equation}
Lastly, we end up with the classical Euler-Lagrange's equation of motion for the complex tachyon fields $T(t), T^*(t)$ given by;
\begin{align}
\ddot{T}&+3H\dot{T}-\frac{1}{a^2}\nabla^2T+\frac{\partial V}{\partial T^*}=0,\\
\ddot{T}^*&+3H\dot{T}^*-\frac{1}{a^2}\nabla^2T^*+\frac{\partial V}{\partial T}=0
\end{align}
with
\begin{align}
H&=\left(\frac{\dot{a}}{a}\right)=\sqrt{\frac{8\pi G}{3}\rho^T},\\
\rho^T&=\abs{\dot{T}}^2+V(T^*T).
\end{align}
Obviously, therefore the remaining ingredient of major concern in the present study of self-gravitating complex tachyon field in the unstable $(D6-\bar{D}6)$ system which is expected to play the role of the inflaton scalar field in the brane inflation scenario while it is (slowly) rolling down its potential is the tachyon potential $V(T^*T)$, which, in the present work, is suggested to be the supergravity analogue constructed by the present author in the unstable $(D6-\bar{D}6)$ pair system represented by a supergravity solution.

\section{The slow roll-over regime}
If, fortunately, the inflation actually takes place, the spacelike hypersurface with the induced 3-metric $h_{ij}$ will get inflated sufficiently, and thus the Gaussian curvature term of this spacelike hypersurface can be neglected. Namely, the Friedmann equation reduces to;
\begin{equation}
\left(\frac{\dot{a}}{a}\right)^2+\frac{k}{a^2}=\frac{8\pi G}{3}\rho^T
\end{equation}
of which the solution is obviously;
\begin{align}
a(t)&=a(0)e^{Ht},\\
H&=\sqrt{\frac{8\pi G}{3}\rho^T}.
\end{align}
Now the conditions that guarantee that this new inflation-type scenario actually takes place can be translated into conditions in the slow roll-over regime as follows

\begin{enumerate} 
\item The existence of vacuum-dominated era, i.e., "inflationary epoch" can be translated into the condition for the inflation scalar field, i.e, the complex tachyon field as;
\begin{equation}
 \rho_m=\frac{1}{2}\lvert\dot{T}\rvert^2+V(T^\ast T)\simeq V(\lvert T\rvert^2),\quad \frac{1}{2}\lvert\dot{T}\rvert^2\ll V(\lvert T\rvert^2)   
 \end{equation}
or into the condition for sufficient inflation, namely sufficient number of e-foldings as:
\begin{align}
N\geq\frac{3H^2}{\lvert V^{''} (\lvert T_{top}\rvert)\rvert}\geq63,\\
Z\equiv\frac{a(t_f)}{a(t_i)}=e^{H\Delta t}\equiv e^{N}.
\end{align}
\item The condition for very flat inflation potential can be translated into the condition for "slow roll-over" as
\begin{align}
\ddot{T}+3H\dot{T}&=-\frac{\partial V(\lvert T\rvert^2)}{\partial T^\ast},\\
\ddot{T}&\ll3H\dot{T},\\
3H\dot{T}&\simeq-\frac{\partial V(\lvert T\rvert^2)}{\partial T^\ast}
\end{align}
or into the conditions on the curvature of the inflaton (i.e., complex tachyon field) potential as; 
\begin{equation}
\abs*{\frac{V'(\vert T\rvert)}{V(\lvert T\rvert)}}\ll\sqrt{48\pi G}
\end{equation}
which is equivalent to; 
\begin{equation}
V^{''}(\abs{T})\ll9H^2.
\end{equation}
\item The condition for acceptable density perturbations that is consistent with the observed "small anisotropy" of 3K "CMBR" (Cosmic Microwave Background Radiation) 
\end{enumerate}

\section{Constraint coming from the density perturbation}
\subsection{(Cosmological) curvature perturbation}
In order to define the cosmological perturbations, one first has to choose a "gauge" (i.e., a coordinate system) which defines a threading and slicing of the spacetime. Now let us denote a generic perturbation by a symbol $g(\vec{x})$. Then at each moment of time (i.e., at each spacelike hypersurface), it is Fourier-expanded in comoving coordinates as;
\begin{equation}
g(\vec{x})=\frac{1}{(2\pi)^{3/2}}\int d^3\vec{k}e^{i\vec{k}\cdot\vec{x}}g(\vec{k})
\end{equation}
with $(k/a)$ being the (proper) wave number.\\ 
Then the spectrum of perturbation is defined by;
\begin{equation}
\langle g(\vec{k})\ g^*(\vec{k'}) \rangle=2\pi^2\frac{1}{k^3}\delta^3(\vec{k}-\vec{k'})P_g(k)
\end{equation}
where the bracket denotes an ensemble average (namely, quantum expectation value if the perturbations originate quantum fluctuations). Then, lastly, the formal expectation value of $g^2(\vec{x})$ at any point in space is;
\begin{equation}
\langle g^2(\vec{x})\rangle=\int^\infty_0d(\ln k) P_g(k)
\end{equation}
and $P^{1/2}_g$ is the typical magnitude of a perturbation in $g(\vec{x})$ with size of order $1/k$. We now get into the specifics of the spatial curvature perturbation.\\\\
\textbf{Spectrum of (curvature) perturbation}

The spectrum of the curvature perturbation during superluminal evolution is found by a standard method. And the result is, well after the horizon exit when $aH=k$, for $a(t)\propto t^{1/3}$
\begin{equation}
P^{1/2}_R\propto\left(\frac{\abs{H}}{M_{pl}}\right)\left(\frac{k}{\alpha\abs{H}}\right)^{3/2}\ln\left(\frac{aH}{k}\right)\\
\end{equation}
where $G=\frac{1}{M^2_{pl}}$ with $G$ being the Newton's constant.

This spectrum is time-independent and corresponds to the spectral index $n=4$. In order to define the spatial curvature perturbation, we need a slicing of spacetime, which is best taken by the orthogonal to comoving worldlines, the "comoving slicing". Then its line element can be written as 
\begin{equation}
dl^2_3=a^2(t)[1+2R]\delta_{ij}dx^idx^j
\end{equation}
or more generally in 4-dim. context,
\begin{equation}
ds^2=N^2(t)dt^2+dl^2_3
\end{equation}
where
\begin{equation}
N^2(t)dt^2=a^2(t)\left\{[1+2\Psi]d\tau^2+[1-2\Phi]\delta_{ij}dx^idx^j\right\} 
\end{equation}
where $d\tau=dt/a$ is the "conformal time" and $R$ defines the curvature perturbation and particularly $R(\Psi,\ \Phi)$ are called Bardeen potential. Finally, the density contrast $\delta\equiv\delta\rho/\rho$ is defined on comoving slices as
\begin{equation}
\delta=-\frac{2}{3}\left(\frac{k}{aH}\right)^2\Phi
\end{equation}
where we selected "comoving slicing" for $R(\Psi,\ \Phi)$.

\subsection{Primordial inflaton(scalar matter field) perturbation}
(Klein-Gordon) field equation for the inflaton scalar field
\begin{align}
\ddot{T}&+3H\dot{T}+\frac{\partial V(T^*T)}{\partial T^*}=0,\\
H&=\sqrt{\frac{8\pi G}{3}\rho^T},\\
\rho^T&=\frac{1}{2}\abs{\dot{T}}^2+V(T^*T).
\end{align}
Note that the inflaton field eq. is equivalent to the Bianchi identity, i.e., energy-momentum conservation; 
\begin{equation}
M^2_{pl}\dot{H}=-\frac{1}{2}\abs{\dot{T}}^2.
\end{equation}
The inflaton (scalar matter) field perturbation starts with the splitting;
\begin{equation}
T=T_0+\delta T
\end{equation}
where $T_0$ and $\delta T$ denotes classical background field and quantum fluctuations, respectively which add up to build the quantum field $T$. The crucial realization regarding this inflaton (scalar) field perturbation is that the quantum fluctuation of this inflaton field $T(x)$ generates the curvature perturbation and then mix with the pure curvature perturbations. Each perturbation $\delta T$ with proper wave number $k/a$ evolves independently and using the conformal time, the quantity
\begin{equation}
u\equiv a\delta T
\end{equation}
  has the same dynamics as a free field living in free flat spacetime, with the wave number just $k$ but with some time-dependent mass squared.
  
Then well before the horizon exit (2nd horizon crossing), at $aH=k$, the mass is negligible, and at the classical level, the perturbation is supposed to vanish (no particles). There is, however, a quantum (or vacuum) fluctuation which, after horizon exit, becomes a classical Gaussian perturbation $\delta T$. Thus when evaluated on spatially-flat slices, this quantum inflaton perturbation determines the (induced) curvature perturbation $R$ as seen by the comoving observers,
\begin{equation}
R=-H\frac{\delta T}{\dot{T}}.
\end{equation}
This quantity is constant during inflation and provided the pressure perturbation is adiabatic, remains constant until horizon entry (1st horizon crossing). And for any given sufficiently-flat inflaton potential, the spectrum of $R$ should be scale-independent (flat) consistently with the (CMB) observation.\\\\
\textbf{Computation of density contrast}
\begin{align}
g&\equiv\frac{\delta \rho}{\rho}=\frac{2}{3}\left(\frac{k}{aH}\right)^2R\le10^{-5}\\
R&=-H\frac{\delta T}{\dot{T}}
\end{align}  
And clearly, the calculation of the density contrast reduces to the task of solving the coupled Friedmann equation and inflaton field equations for $a(t),\ T(t)'$ to determine $H$ and $R$.\\\\
\textbf{Coupled Friedmann equation and inflaton field equation}
\begin{align}
\left(\frac{\dot{a}}{a}\right)^2+\frac{k}{a^2}&=\frac{8\pi G}{3}\rho^T\\
\ddot{T}+3H\dot{T}&=\frac{-\partial V(\abs{T}^2)}{\partial T^*}
\end{align}
where
\begin{equation}
H=\sqrt{\frac{8\pi G}{3}\rho^T},\ \rho^T=\frac{1}{2}\dot{T}^2+V(\abs{T}^2)
\end{equation}
\textbf{Solution ansatz power-law inflation}
\begin{align}
a&\propto t^p\  (p>1)\\ 
H&=\left(\frac{\dot{a}}{a}\right)=\frac{p}{t}\\
H&=\sqrt{\frac{8\pi G}{3}\rho_m}\propto\sqrt{\frac{8\pi G}{3}\frac{1}{2}\dot{T}^2}\\
\frac{1}{2}\dot{T}^2&\propto\frac{H^2}{G}=M^2_{pl}H^2\\
d\tau&=\frac{dt}{a},\ \tau=\frac{t}{a}\propto t^{(1-p)}\\
aH&=pt^{p-1}=\frac{p}{\tau}\quad \text{or}\quad \tau=\frac{p}{a+1} 
\end{align}
where $\tau$ denotes a conformal time.\\\\
\textbf{Spectrum of (curvature) perturbation}
\begin{equation}
P^\frac{1}{2}_R\propto\left(\frac{H}{M_{pl}}\right)\left(\frac{k}{aH}\right)^{\frac{1}{1-p}}
\end{equation}
This spectrum corresponds to the spectral index $n=1+\frac{2}{1-p}$.\\\\
\textbf{Computation of perturbation including the metric perturbation}

Below, we summarize the calculation of density perturbation and hence the spectrum of perturbation. Indeed, in the case of slow-roll inflation, the metric perturbation is negligible compared with the inflaton (scalar matter) field perturbation. In general, however, they are of the same order and thus the metric perturbation should be considered as well.\\\\
\textbf{Computation of perturbation ignoring the metric perturbation}

Although we shall eventually include the metric perturbation we first start with a rough argument. The inflaton field perturbation $\delta T$ in the splitting;
\begin{equation}
T=T_0+\delta T
\end{equation}
would satisfy, neglecting the metric perturbation, the inflaton (scalar) field equation
\begin{equation}
\delta\ddot{T}+\left[\left(\frac{k}{a}\right)^2+V^{''}\right]\delta T=0
\end{equation}
which can be rewritten as;
\begin{equation}
\delta\ddot{T}+\left[\frac{k^2}{a^2}-\frac{2}{t^2}\right]\delta T=0.
\end{equation}
Namely, this is the fundamental equation that the inflaton (scalar field) perturbation satisfies. Note that the amplification (inflation) of the quantum fluctuation (i.e., of the inflaton perturbation) takes place around the epoch $\left(\frac{k}{a}\right)^2=V^{''}(T)$. Indeed, well before this epoch, the quantum fluctuation is that of a massless free field in flat spacetime, which on dimensional grounds, has the spectrum $P_T\sim\left(\frac{k}{a}\right)^2$. Next, for the scale of an estimate, we just assume that this expression remains valid at the epoch of amplification $\left(\frac{k}{a}\right)^2=V^{''}(T)$. Now, well-after this epoch, the negative-mass squared $V^{''}(T)$ dominates and $\delta T$ increases. In fact, $\delta T\propto\frac{1}{t}\propto\dot{T}$ so that $\frac{P_T}{\dot{T}^2}\sim\abs{\frac{V^{''}}{V}}=\frac{2}{pM^2_{pl}}$ which gives $P_R\sim\left(\frac{2}{p}\right)\left(\frac{H}{M_{pl}}\right)^2$. Next, in order to obtain a more precise result, we employ the quantity, $u\equiv a\delta T$ and the conformal time, $\delta\tau=\delta t/a$. Then the fundamental equation that the inflaton perturbation satisfies given above becomes
\begin{equation}
\frac{d^2u}{d\tau^2}+\left[k^2-\frac{2}{\tau^2}\right]u=0.
\end{equation}
We are now ready to summarize the computation of perturbation including the metric perturbation that we were originally aimed at. Generally when the inflaton (scalar matter) field perturbation and the metric perturbation are of the same order, the fundamental equation that the generic perturbations satisfy is replaced by
\begin{equation}
\frac{d^2u}{d\tau^2}+\left[k^2-\frac{1}{z}\frac{d^2z}{d\tau^2}\right]u=0
\end{equation}
where $z\equiv (a\dot{T})/H$ is newly introduced.

The consistency of this fundamental equation for both cases with and without the metric perturbation can be checked immediately as follows;
\begin{align} 
z&=\frac{a\dot{T}}{H}=a\sqrt{2}M_{pl}\\ 
&=\sqrt{2}M_{pl}t^p=\sqrt{2}M_{pl}\tau^{\frac{p}{1-p}},\\
\frac{d^2z}{d\tau^2}&=\sqrt{2}M_{pl}\frac{-p(1-2p)}{(1-p)^2}\tau^{\frac{p}{1-p}-2},\\
\frac{1}{z}\frac{d^2z}{d\tau^2}&=\frac{-p(1-2p)}{(1-p)^2}\tau^{-2},\\
\frac{d^2u}{d\tau^2}&+\left[k^2+\frac{p(1-2p)}{(1-p)^2}\tau^{-2}\right]u=0.
\end{align}
It is now manifest that the power-law inflation corresponds to $p>1$ giving a negative mass-squared which amplifies the quantum fluctuation (namely, the inflaton perturbation).

The spectrum of curvature perturbation well after the horizon exit, then, is
\begin{equation}
P^{1/2}_R=\sqrt{\frac{p}{2}}\left(\frac{H}{M_{pl}}\right)\left(\frac{k}{aH}\right)^{1/(1-p)} 
\end{equation}
which is time-independent and corresponds to the spectral index $n=1+\frac{2}{1-p}$.
Note first that the eqs.(6.35) and (6.36) are essentially the same considering the definition for $z$ given in eq.(6.33). Before we leave this discussion, we briefly provide the analytic, closed form solution for $u=u(\tau)$ and $z=z(\tau)$ as a solution for the eq.(6.37) and (6.35) respectively which would represent how the spacetime geometry and the matter field(tachyon) would be affected by the inflation. First, the eq.(6.35) is easily integrated to yield $z(\tau)=\sqrt{2}M_{pl}\tau^{\frac{p}{1-p}}$. Next, we turn to the equation (6.37).

According to the definition of "Whittaker functions" \cite{5}. It is the solution of
\begin{equation}
\frac{d^2W}{d\tau^2}+\left(-\frac{1}{4}+\frac{\lambda}{\tau}+\frac{\frac{1}{4}-\mu^2}{\tau^2}\right)W=0
\end{equation}
and generally this differential equation admits the following two linearly independent solutions;
\begin{align}
M_{\lambda, \mu}(\tau)&=\tau^{\mu+\frac{1}{2}}e^{-\frac{\tau}{2}}\Phi(\mu-\lambda+\frac{1}{2}, 2\mu+1;\tau),\\
M_{\lambda, -\mu}(\tau)&=\tau^{-\mu+\frac{1}{2}}e^{-\frac{\tau}{2}}\Phi(-\mu-\lambda+\frac{1}{2}, -2\mu+1;\tau)
\end{align}
where $\Phi$ denotes a degenerate hypergeometric function \cite{5}.

As a result, the solution for eq.(6.37) can be identified with the Whittaker function for the particular case;
\begin{align}
k^2&=-\frac{1}{4},\\
\lambda&=0,\\
\mu^2&=\frac{1}{4}-\frac{p(1-2p)}{(1-p)^2}.
\end{align}

To summarize, in order for this open string tachyon$=$stringy inflaton scenario to be successful, the bottomline condition, namely, enough inflation (63-efolding) should be supplemented by not too much density/perturbations, $\frac{\delta\rho}{\rho}\le10^{-5}$. In principle, the "enough inflation" condition appears to be safely met by eq.(4.16). In order to secure the other condition, "not too much density/perturbations", however, we need to solve the Euler-Lagrange's eq. of motion for the tachyon field $T(t)$, which, is coupled to the Friedmann eq. for $a(t)$. This is, indeed, rather a delicate, complex nature to tackle that may demand a numerical approach. In the present work, therefore, we demonstrate the overall algorithm to achieve this goal but still leave the actual computations for future study.

\begin{figure}[h!]
\includegraphics[width=10cm, angle=0]{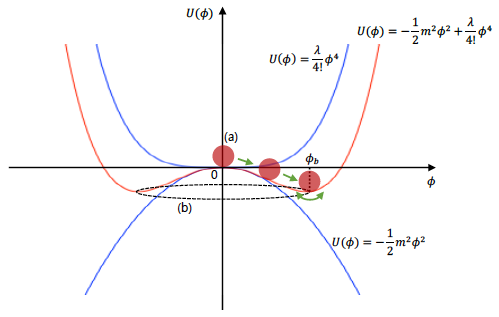}
\caption{SSB-inspired (Mexican hat) potential of a neutral(uncharged) scalar field theory
 rolling down from the false vacuum(a) to the true vacuum(b) represents the (open) string tachyon condensation.}
\label{TCfigure}
\end{figure}
\newpage
\noindent Uncharged (neutral) scalar field theory (with the tachyon mode being present)
\begin{equation}
\mathcal{L}=T(\phi)-U(\phi)\ \ \  \text{(Lagrangian density)}
\end{equation}
with
\begin{align}
T(\phi)&=-\frac{1}{2}g^{\mu\nu}\partial_\mu\phi\partial_\nu\phi=\frac{1}{2}(\partial_t\phi)^2-\frac{1}{2}(\nabla\phi)^2,\\
U(\phi)&=-\frac{1}{2}m^2\phi^2+\frac{\lambda}{4!}\phi^4 
\end{align}
where the potential energy $U(\phi)$ is the SSB inspired (Mexican hat) potential which is known to be renormalizable in ($d=4$).
\begin{equation}
m^2=\left.-\frac{\partial^2U}{\partial\phi^2}\right|_{\phi=0}=
\begin{cases}
>0\ (\text{physically-acceptable mode})
\\<0\ (\text{physically-unacceptable tachyon mode})
\end{cases}
\end{equation}
\paragraph{Note}
Given a (super)string theory, if it admits (open string) tachyon(s)... How one gets rid of them? tachyon condensation! Build a tachyon potential that admits a lowest-lying mode which turns the existing tachyon mode into a physically-acceptable $m^2>0$ mode.

\section{Criterion for successful stringy inflation}
The bottomline is to demand enough inflation, $Z=e^N$ (eq.(5.6)),$\ N\geq63$ (63-efolding) to resolve all the cosmological problems:
\begin{itemize}
\item[I.] Horizon problem
\item[II.] Flatness problem
\item[III]. sufficient Reheating
\item[IV.] not too much density perturbations to be consistent with CMBR data \cite{6, 7, 8} $\frac{\delta\rho}{\rho}\leq10^{-5}$ 
\end{itemize}
In order to address these issues, in principle, one has to solve the coupled Einstein equation for $a(t)$, Klein-Gordon equation for $T(t)=\phi(t)$ simultaneously.

(A) In the present work, however, instead of solving numerically these coupled non-linear differential equations, we constructed analytic, closed form solution for $(u(\tau), z(\tau)) \sim (a(\tau), T(\tau))$ such that successful stringy inflation guided by the 4-conditions I, II, III, IV could be satisfied.

(B) and practically speaking, solving these coupled, non-linear 2nd differential equations usually demands numerical integration. The other avenue, namely, the numerical integration of the differential equations to address this issues: I, II, III, IV, therefore, still remains to be studied and hence left for interested community of researchers.

\section{Concluding remarks}
It has been known for some time that the present superstring context involves AdS vacua but not dS vacua where AdS and dS denotes negative and positive vacuum energy (or the cosmological constant) respectively. In the mean time, since the advent of inflationary universe paradigm, the presence of inflation in the early universe actually has been established as a standard cosmological model scenario. As such, since the current superstring theory indeed has attained the status of the reliable theory of quantum gravity at even earlier universe epoch, in order to reconcile the current superstring theory with the established early universe cosmology, therefore, it is essential to implement the inflation in the context of the current superstring theory. Along this line, the early works dubbed, KKLT and KKLMMT have played the pioneering role. In the present work, we have demonstrated that amazingly it is the open string tachyon which is the fully qualified stringy inflaton. Put differently, it is the open string tachyon condensation process that is the long-seeking stringy inflation in disguise. Once again, it is indeed amazing that unlike ones plain conjecture, the unwelcome tachyon in open string theory turns out to be the long-seeking and hence mysterious, missing stringy inflaton.



\end{document}